# A lithium-ion battery based on a graphene nanoflakes ink anode and a lithium iron phosphate cathode

Jusef Hassoun[1§], Francesco Bonaccorso[2,3§*], Marco Agostini[1], Marco Angelucci[4], Maria Grazia Betti[4], Roberto Cingolani[5], Mauro Gemmi[6], Carlo Mariani[4], Stefania Panero[1], Vittorio Pellegrini[5,3] and Bruno Scrosati[5]

[1] Department of Chemistry, University Sapienza of Rome, I-00185 Rome, Italy
[2] CNR-Istituto per i processi Chimico-Fisici, I-98158 Messina, Italy
[3] NEST, Istituto Nanoscienze-CNR, and Scuola Normale Superiore, I-56127 Pisa, Italy
[4] Department of Physics, University Sapienza of Rome, I-00185 Rome, Italy
[5] Istituto Italiano di Tecnologia, Graphene Labs, I-16163 Genova, Italy
[6] Center for Nanotechnology Innovation@NEST, Istituto Italiano di Tecnologia, I-56127 Pisa, Italy.

§ These authors contributed equally to this work.
*Corresponding author: bonaccorso@me.cnr.it

Li-ion rechargeable batteries have enabled the wireless revolution transforming global communication. Future challenges, however, demands distributed energy supply at a level that is not feasible with the current energy-storage technology. New materials, capable of providing higher energy density are needed. Here we report a new class of lithium-ion batteries based on a graphene ink anode and a lithium iron phosphate cathode. By carefully balancing the cell composition and suppressing the initial irreversible capacity of the anode, we demonstrate an optimal battery performance in terms of specific capacity, *i.e.* 165 mAhg$^{-1}$, estimated energy density of about 190 Whkg$^{-1}$ and life, with a stable operation for over 80 charge-discharge cycles. We link these unique properties to the graphene nanoflake anode displaying crystalline order and high uptake of lithium at the edges, as well as to its structural and morphological optimization in relation to the overall battery composition. Our approach, compatible with any printing technologies, is cheap and scalable and opens up new opportunities for the development of high-capacity Li-ion batteries.

The development of next-generation portable electronics and electric vehicles[1] is inherently connected with advances in energy storage devices such as batteries[2] and supercapacitors[3]. In particular, Li-ion batteries[2,4,5] are currently dominating the market for portable electronic devices (*e.g.* laptops, mobile phones, etc.)[6]. In their most common configuration, these batteries are composed by an intercalated Li compound cathode (*e.g.* LiCoO$_2$ or LiFePO$_4$), a graphitic anode and an electrolyte with achieved energy densities of the order of ~120-150 Wh kg$^{-1}$ [5]. However, Li-ion batteries do not meet criteria performances, in terms of cost (~120US$ per kWh)[1], charge/discharge rate (full charge in ~30mins for 200 km electric car battery)[1], measured energy density (>300 Wh kg$^{-1}$)[1] and safety, needed for all-electric vehicles development as well as for the storage of electrical energy converted by wind and/or solar power, making present Li-ion technology rather problematic as a lone energy source[7].

Energy and power capability of a lithium-ion battery critically depends on the rate at which the Li$^+$ ions and electrons can migrate from the electrolyte and electrode support, respectively, into the active electrode material as well as on the gravimetric capacity to store Li ions (*i.e.* the weight percentage of stored Li per gram of battery weight)[4,5,6]. A critical issue is the low theoretical specific capacity (*i.e.* the total ampere-hours -Ah- available when the battery is discharged at a certain discharge current, per unit weight) of graphite anodes (372mAh g$^{-1}$ [5]). For this reason a large fraction of current research is focusing on alternative anode materials such as Si (4200 mAhg$^-$

$^1)^8$, Sn (994 mAhg$^{-1}$)$^9$ and SnO$_2$ (782 mAhg$^{-1}$)$^{10}$. However, their application has been mostly limited by their poor cycling (*i.e.* the number of charge/discharge cycles before the specific capacity falls below 60% of nominal value) caused by large volume changes (100-300% with respect to the initial volume)$^{11}$ during the repeated alloying and de-alloying process with Li$^+$.

Graphene, thanks to its large surface to mass ratio (SSA) exceeding 2600 m$^2$g$^{-1}$ [12], high electrical conductivity ($\sigma$)$^{13}$, high mechanical strength$^{14}$, with the added value of mass production$^{15}$, is a promising material for electrodes in Li-ion batteries$^{16,17}$. While single layer graphene (SLG), grown via chemical vapour deposition (CVD)$^{18}$, has a limited capability of uptaking Li ions (5% surface coverage) due to repulsion forces between Li$^+$ at both sides of the graphene layer$^{19}$, large efforts have been devoted to the exploitation of chemically modified graphene (CMG) such as graphene oxide (GO) and reduced GO (RGO), both at the anode$^{16,20,21,22,23,24,25,26}$ and cathode$^{27,28,29,30}$. However, although CMG can be produced in large quantities they suffer from limited $\sigma^{31}$ and diffusion of Li ions. To date, the best anodes with CMG have reached specific capacity of ~1200 mAhg$^{-1}$ at 100 mAg$^{-1}$ current rate in half cell$^{20}$ and ~100 mAhg$^{-1}$ at 29 mAg$^{-1}$ current rate when assembled in full battery$^{32}$.

Graphene nanoflakes, obtained from the exfoliation of pristine graphite, represent an ideal yet unexplored material platform for battery electrodes. Indeed they possess high crystallinity$^{16}$ key for fast electron transport to the electrode support$^5$. Moreover, graphene nanoflakes having small (<100 nm) lateral size offer a large edge to bulk ratio of carbon (C) atoms. This is crucial since edges are considered very active sites for Li$^+$ storage$^{33}$, providing much stronger (up to 50%) binding energies for Li$^+$ (1.70-2.27 eV) with respect to graphene basal plane (1.55 eV)$^{33}$. Additionally, theoretical calculations, for graphene nanoribbons (GNRs), predict that edges also provide decreased energy barriers for Li diffusion$^{33}$, up to 0.15 eV smaller than those in graphene basal plane$^{33}$. Finally such graphene nanoflakes can be produced by simple methods in the form of inks$^{34}$ and deposited by conventional printing processes$^{15}$ to form electrodes of the desired shape and functionality.

Anodes based on such nanostructured systems could in principle reach higher gravimetric and specific capacity than graphene (*e.g.* theoretical specific capacity of ~740mAhg$^{-1}$[35]) and CMG. Despite all these promising properties, however, such graphene-based material system has not been used so far as anode in batteries. One additional challenge is the assembly of such high-capacity anode materials in a balanced battery exploiting conventional cathodes (*e.g.* LiCoO$_2$ or LiFePO$_4$)$^{4,5,6}$ with specific capacity of ~150 mAhg$^{-1}$ [5]. Indeed, a correct anode to cathode balance (in terms of specific capacity and weight) is a key requirement$^5$ to assuring proper battery performances in terms of cycle life and capacity stability.

Here we successfully produce inks of graphene nanoflakes with controlled morphological properties (lateral size and thickness) obtained by the combination of liquid phase exfoliation (LPE) of graphite$^{36}$ and sorting strategy in centrifugal field$^{15}$. By a combination of Raman and X-ray photoelectron (XPS) spectroscopies we assess the structural quality of the graphene nanoflakes, unraveling the mechanism for Li ions uptake. We show that electrodes based on Cu-supported graphene nanoflakes ink can reach specific capacities of ~1,500 mAhg$^{-1}$ at a current rate of 100 mAg$^{-1}$ over 150 cycles, well beyond the theoretical specific capacity of graphene$^{35}$. We also demonstrate the first Li-ion battery based on Cu-supported graphene nanoflakes anode and a lithium iron phosphate cathode. This new type of battery displays excellent performances, due to optimal cathode/anode balance and suppression of the initial irreversible capacity of the anode, reaching a reversible capacity of 165 mAhg$^{-1}$ and operating for over 80 charge-discharge cycles at 1C rate, equivalent to battery charge/discharge to maximum capacity in 1 hour, with a Coloumbic efficiency approaching 100%.

The use of inks of graphene nanoflakes with controlled geometry instead of other graphene-based nanomaterials represents an innovative direction for the realization of a new class of electrical energy-storage devices at greatly reduced cost and with greatly-improved specific capacity with respect to the current technology[4,5,6,7]. The inherent scalability of the production process, and the wide possibilities for further optimization of battery's performances, suggest that the approach here demonstrated might impact available technological solutions for energy storage.

**Graphene nanoflake inks**

Liquid phase exfoliation produces small flakes[37], with a large number of active edge sites that, per unit mass, are more than those in both graphite and large graphene flakes.

We exfoliate graphite by chemical wet dispersion[34,36], see Fig.1a. In order to obtain graphitic flakes with small lateral size, after the exfoliation process, the as-prepared dispersion is purified via ultracentrifugation[15, 34, 36] exploiting the sedimentation-based separation (SBS)[15], see Fig. 1b. This separation technique separates graphitic flakes (particles in general[38]) into fractions on the basis of their sedimentation rate, which determines how graphitic flakes (particles) in dispersion sediment out of the fluid in which they are dispersed, in response to a centrifugal force acting on them[15]. Thus, SBS of graphitic flakes ensures a separation based on their mass and shape[15]. This step enables the control of the lateral dimension of the nanoflakes. For anode optimization, we believe it is crucial to produce small flakes with dimension below 100 nm in order to enhance the number of edge sites per unit mass still preserving their high σ.

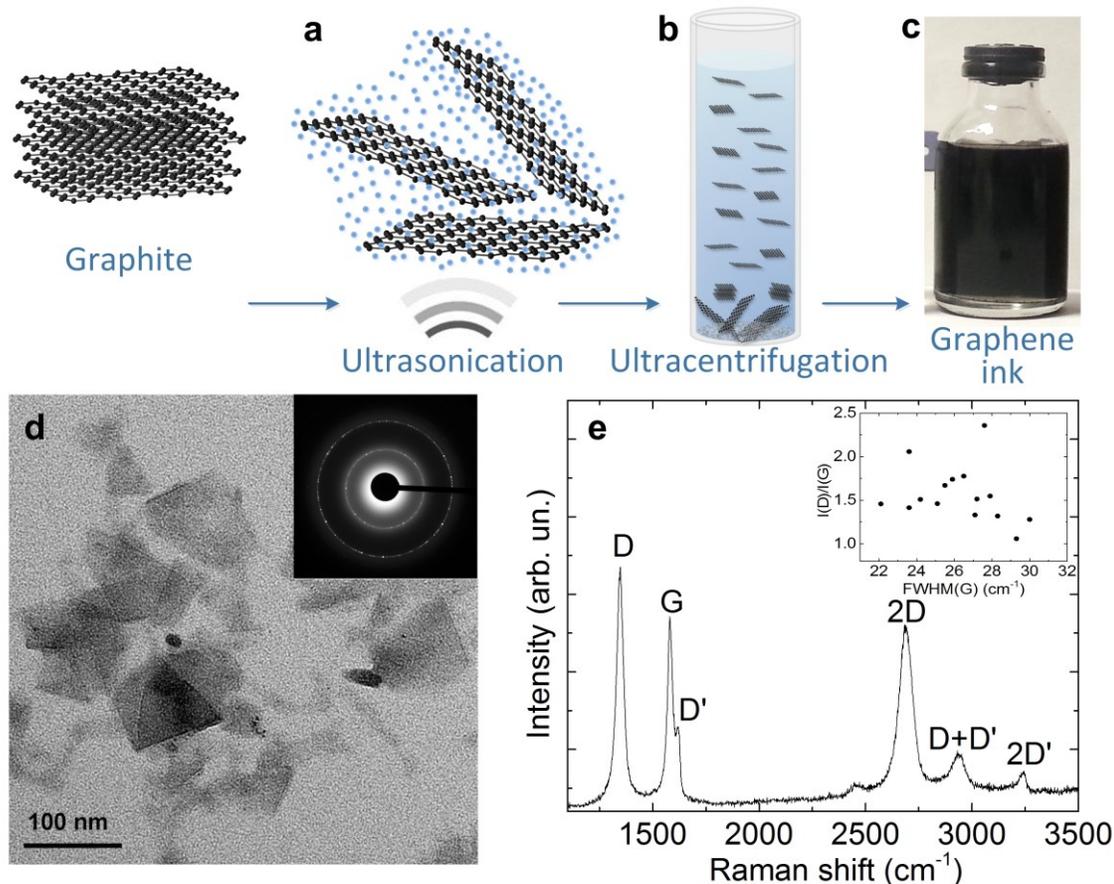

Figure 1. Production and characterization of graphene nanoflakes ink. **a**, Liquid phase exfoliation of graphite via ultrasonication[36,34]. Graphite is dispersed in solvent, schematically represented by blue dots, and exfoliated with the aid of ultrasound. **b**, Small lateral size flakes are separated from thick and large flakes via ultracentrifugation. **c**, Photograph of graphene ink. **d**, Bright field TEM image of graphene flakes at higher magnification. Inset: Electron diffraction pattern collected on an area of 2μm in diameter. The 10-10 and 11-20 polycrystalline diffraction rings of graphene are clearly visible. The rings are formed by strong spots corresponding to the larger flakes and a background of weaker unresolved spots associated to smaller flakes. **e**, Raman spectrum measured at 532 nm laser excitation wavelength for a representative flake obtained by SBS. Inset: I(D)/I(G) as a function of full-width half maximum of the G peak -FWHM(G)- measured on flakes deposited on Si/SiO$_2$.

To this end, we ultracentrifuged the dispersion at ~77000g for 30mins, a value ~5 times higher with respect to the one used in Ref. 34, where the resulting flakes had a lateral size of 300-1000 nm. A detailed explanation of the exfoliation and sorting procedures is given in appendix A. Using the optical absorption coefficient (α) of 1390 $Lg^{-1}m^{-1}$ at 660 nm[36,34] we estimate, via optical absorption spectroscopy, a concentration of graphitic material in the resulting ink of ~40 mg/l, see Fig. 6 in appendix A. The graphene ink, Fig. 1c, mostly contains flakes having variable dimensions mostly in the range ~30–100 nm as shown by the TEM analysis, reported in Fig. 1d. The inset to Fig. 1c shows the normal-incidence electron diffraction pattern collected on flake aggregates[39]. The diffraction pattern (see Fig. 7 in appendix B) shows polycrystalline rings demonstrating that the flakes are crystalline. All the rings can be indexed as h,k,-h-k,0 reflections of an hexagonal lattice as expected for graphite/graphene flakes. Figure 1e plots a typical Raman spectrum of the flakes deposited on $Si/SiO_2$. Besides the G and 2D peaks, fingerprints of graphene[39], this spectrum shows significant D and D' intensities and the combination mode D+D' (see appendix B for a more complete discussion of the peak assignment and Raman analysis). Statistical analysis (Fig. 8 in appendix B) demonstrates that the ink contains a combination of SLG and few-layer graphene (FLG) flakes. The high intensity ratio I(D)/I(G)~1.50 (see Fig. 8d) is attributed to the edges of our nanometer flakes[40], rather than to the presence of a large amount of structural defects within the flakes. This is confirmed by the lack of a clear correlation between I(D)/I(G) and the FWHM(G) (see inset to Fig. 1e).

**Graphene nanoflake anode**

The anode electrode is formed by drop casting the graphene nanoflake ink (see Fig. 2a) at 140°C in ambient condition on a polycrystalline Cu support shown in Fig. 2b. In order to clean the graphene flakes from any solvent contamination, the sample is then annealed at 400 °C under ultra-high vacuum (UHV) thus avoiding any reducing gaseous ambient[41]. The scanning electron microscopy (SEM) image of the thermally-treated sample (Fig. 2c) shows the as-deposited graphene nanoflakes with lateral sizes of the order of tens of nm.

Despite the small lateral dimensions of the nanoflakes, the electrode displays excellent electrical properties, fundamental for electron transport to the external circuit[5]. The measured value (average on 10 measurements) of sheet resistance ($Rs$) of the electrode is ~ 0.5 Ω/□.

Figure 2d compares a typical Raman spectrum of graphene nanoflakes deposited from the ink, with the one measured on the electrode. The Raman spectrum of the electrode has Pos(2D) upshifted (~10$cm^{-1}$) and FWHM(2D) slightly broader (4$cm^{-1}$) with respect to that of graphene flakes in the ink, see Fig. 9 in appendix C. However, the 2D peak still shows a Lorentzian lineshape and in any case distinctly different from that of graphite[42]. This implies that the flakes are SLG or they are electronically almost decoupled in the case of FLG and behave, to a first approximation, like a collection of SLGs[37]. Moreover, I(D)/I(G) and FWHM(G) are not correlated, again showing the lack of large amounts of defects within the flakes composing the electrode, see Fig. 2e.

To gain a better understanding of the quality of the graphene nanoflake electrode and investigate the mechanism of Li uptake, we perform XPS analysis focussing on the C 1s core-level[43]. The C 1s core-level reported in Fig. 2 (f) is a well-defined peak at 284.1 eV binding energy (BE), associated to high-purity graphene[44] with only a negligible high BE tail due to carbon-oxygen bonds[45], *i.e.* C-O, C=O and O-C=O .

The graphene nanoflakes electrode is exposed to Li in controlled evaporation UHV conditions ($10^{-10}$ mbar). Upon exposure to Li, the C 1s XPS signal associated to graphene[44] upshifts by 0.4 eV towards higher BE (see Fig. 10b in appendix C) reaching a value of 284.5 eV after 60' of Li exposure. This energy upshift, coupled with an intensity reduction of the C 1s XPS signal with respect to the starting sample, is related to the bonding of Li ionized ions to C. After 60' of Li exposure, the C 1s peak does not display further energy shift, suggesting the achievement of Li saturation for the graphene nanoflakes. The Li 1s core level at saturation coverage is reported in Fig. 2g (lower panel). The photo-excitation cross-section of 1s core levels of Li and C atoms is 0.79 × $10^{-3}$ Mbarn and 13.67 × $10^{-3}$ Mbarn, respectively[46]. This implies a stoichiometry corresponding to one Li ion per two C atoms of the graphene flakes. In addition, such optimal stoichiometry condition is stable and it can be linked to the Li ion uptake both on graphene nanoflakes basal plane and at their edges. To date, high Li ions concentration has only been achieved in super-dense Li-graphite intercalated compounds[47,48], obtained by high pressure synthesis. However, theory has predicted[48] that $LiC_2$ stoichiometry in lithium-graphite intercalated compounds is metastable at ambient pressure[48], which has been also experiment confirmed[48]. Other carbon nanostructures, such as CMG[20,21,22,23,24], carbon nanotubes (CNTs)[49] and disordered carbon structures[50], although have shown higher Li ions uptake than graphite[51], have not matched the $LiC_2$ stoichiometry. Thus the results here reported demonstrate that graphene nanoflakes are most suitable as a nanostructured carbon-based material for Li ions uptake.

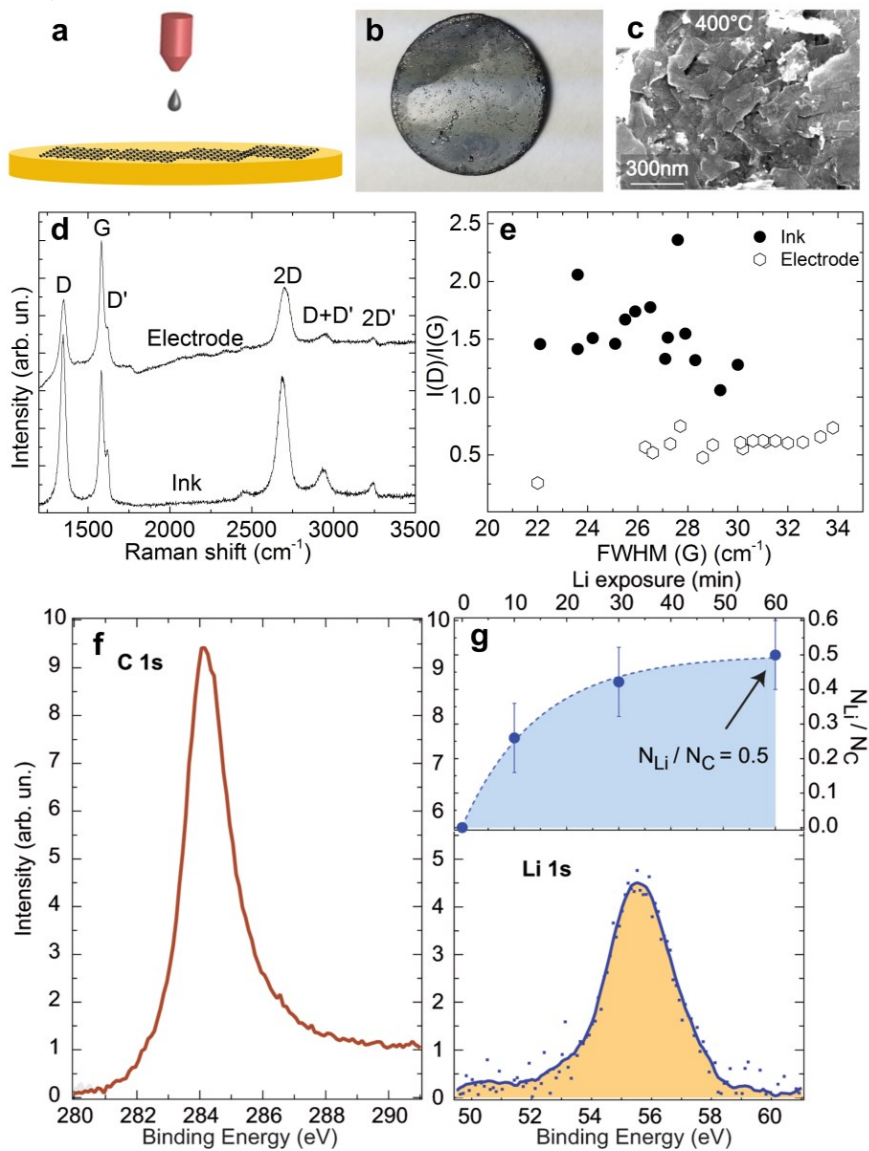

**Figure 2. Preparation and characterization of graphene nanoflakes Cu-supported electrode. a**, Drop casting of graphene ink on Cu substrate. **b**, Photograph of the graphene-Cu supported electrode. **c**, SEM image taken on graphene/Cu after thermal annealing at 400 °C in UHV. **d**, Representative Raman spectrum of flakes dispersed in the ink, compared with the spectrum measured on the Cu-supported graphene electrode. **e**, I(D)/I(G) as a function of FWHM(G) for the ink (filled dots) and the electrode (open dots). **f**, XPS spectral density of the C 1s core-level. **g**, XPS spectral density of the L 1s core-level (lower panel) and estimated number of Li atoms per C atom, obtained through the weighted intensity ratio of the core-levels.

Motivated by the results reported above we assess the electrochemical performance of a graphene nanoflake electrode. Fig. 3a shows the voltage profile of the first discharge performed at a current of 700 mA g$^{-1}$ of the annealed Cu-supported graphene electrode in a lithium cell. The extremely high specific capacity (~7500 mAhg$^{-1}$) observed here can only be explained by assuming that this first process involves, in addition to the lithium uptake process on the graphene flakes, also, and mainly, side reactions such as those associated with the decomposition of the electrolyte[52], which cause the formation of a passivating film or solid electrolyte interphase (SEI) on the surface of the carbon electrode[52]. The occurrence of side reactions, irreversible in nature[52], is typically observed in lithium-based cells[5,52] and associated with the high SSA of the electrodes[52]. The high irreversible specific capacity of the Cu-supported graphene nanoflakes electrode is also favoured by the reactivity of the edges[33]. Fig. 3b shows that the following charge/discharge cycles evolve with an initial capacity fading and stabilization at about 750 mAh g$^{-1}$ (steady state capacity, see also corresponding voltage profile in the inset of Fig.3a). The high initial irreversible capacity is an undesired event in view of practical applications and this issue is here addressed by ex-situ lithiation processes, carried out by directly contacting a lithium metal foil, wet by the electrolyte, with the Cu-supported graphene nanoflakes electrode.

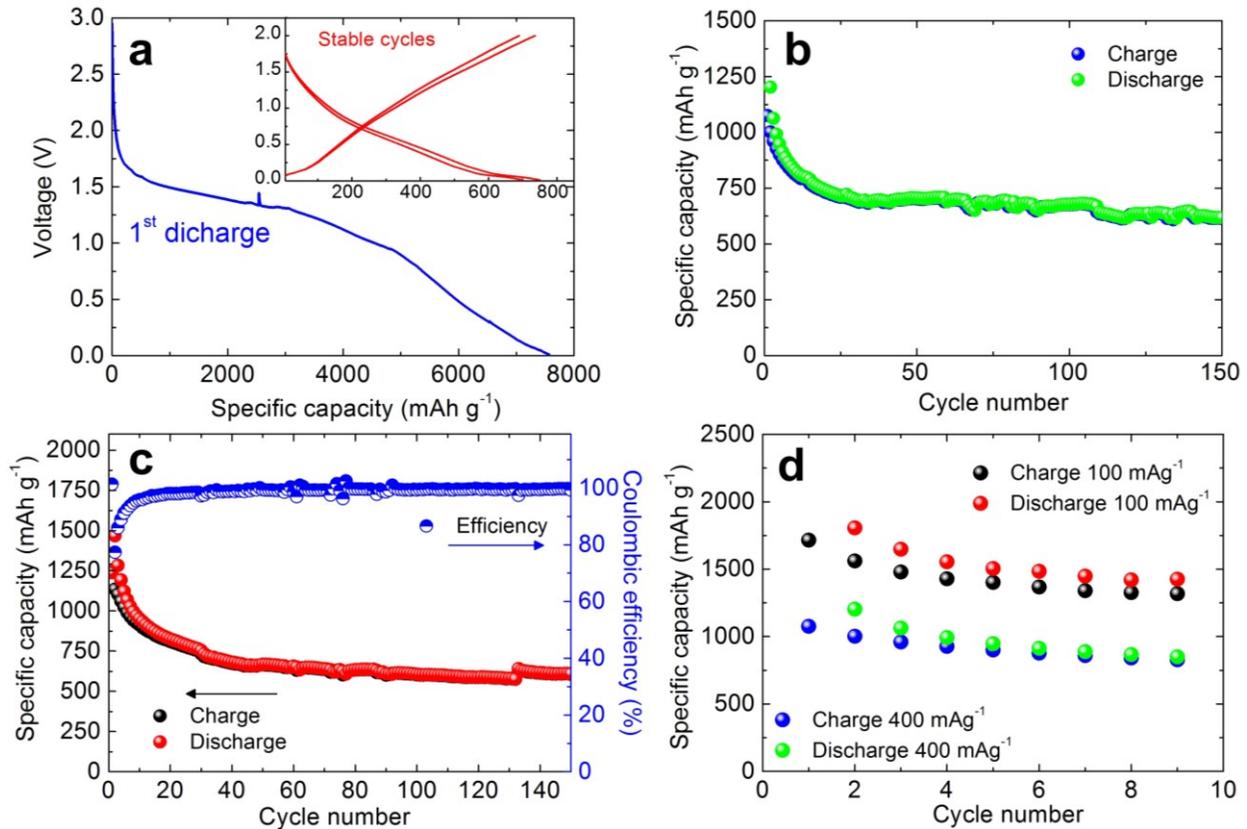

**Figure 3: Electrochemical characterization of the Cu-supported graphene electrode in a lithium cell. a**, Voltage profile of the first discharge (Inset: reversible steady-state profile at the 50th cycle) and **b**, cycling response of the cell during the following cycles. Rate: 700 mA g$^{-1}$, voltage limits 0.01V-2V. **c,** Prolonged cycling (black/red dots) following an ex-situ lithiation procedure and corresponding Coulombic efficiency (blues dots). Rate: 700 mA g$^{-1}$, voltage limits 0.01V-2V. **d**, Specific capacity versus cycle number at rates of 100 and 400 mAg$^{-1}$.

Fig. 3c reports the cycling response of the pre-treated (ex-situ lithiated) electrode, at charge-discharge rate of 700 mAg$^{-1}$. The irreversible capacity is practically vanished, with a first cycle Coulombic efficiency approaching 100%. A slight decrease of the Coulombic efficiency is observed during the following few cycles (~10), most likely associated to a residual SEI film formation process. The subsequent increase and stabilization of the Coulombic efficiency (after >10 cycles) at ~100% confirms the success of the activation procedure, leading to a reversible specific capacity of ~650 mAh g$^{-1}$ upon 150 charge/discharge cycles. Fig. 3c shows the rate capability of the Cu-supported graphene nanoflakes electrode at two charge-discharge rates, at 100 mA g$^{-1}$ and 400 mA g$^{-1}$, respectively. When cycled at low rate (100 mAg$^{-1}$), the electrode delivers a very high reversible capacity *i.e.*, higher than 1,500 mAhg$^{-1}$, almost doubling the theoretical value for graphene[35]. This high reversible capacity value can be accounted for considering that the relatively low current rate of 100mAg$^{-1}$ may provide the "ideal" condition for the uptake of a large amount of Li ions throughout the randomly stacked, (see Fig. 2c) graphene nanoflakes. These remarkable values outperform those obtained with other C-based nanostructures such as GNRs[53,54,55] having a reversible capacity of ~825 mAhg$^{-1}$ at a current density of 100 mAg$^{-1}$ [53]. The Cu-supported graphene nanoflakes electrode has also long cycle life (over 150 charge-discharge cycles) and Coloumbic efficiency of ~100%, Fig. 3d.

**Graphene nanoflake full battery**

In the following we focus on the realization of a full Li-ion battery (Fig. 4a), by coupling the Cu-supported graphene nanoflake anode with a lithium iron phosphate, LiFePO$_4$, cathode, commonly used in commercial batteries[5,6]. In designing the battery, it is of paramount importance not only the electro-chemical performances (*i.e.* σ, specific capacity, cycle life, Coulumbic efficiency, etc.)[5], but also the optimal balance of cathode and anode electrodes both in term of weight and electro-chemical properties. The parameters we used in trying to optimize the cathode/anode balancing are reported in methods. Figure 4b compares the reversible voltage profile versus Li of the Cu-supported graphene nanoflake anode (black line) and of the LiFePO$_4$ cathode (blue line). The anode operates reversibly with continuous, plateau-free[32] charge-discharge curves with a specific capacity of about 700 mAhg$^{-1}$ at an average voltage value of about 1.3 V, while the LiFePO$_4$ cathode cycles with reversible capacity of 165 mAhg$^{-1}$ at a voltage value of 3.5 V vs. Li with a flat plateau, typical of the two phases reaction of lithium-iron olivine[5]. The battery is thus characterized by a slight excess of anode capacity, achieved taking into account a 1:4 mass ratio of graphene in respect to LiFePO$_4$, value calculated to account for the difference in specific capacity of the two electrodes, to finally approach the cell capacity balance to the 1:1 ratio needed for the optimization of the cell.

Figure 4c reports the trend of the full-cell voltage profile demonstrating a very stable behavior. The cell operates with voltage at around 2.3 V and the voltage profile is the combination of the flat voltage of the LiFePO$_4$ cathode (Fig. 4b blues curves) and the sloppy shape of the graphene nanoflakes anode (Fig. 4b black curves). The reversible capacity of the battery is as high as 165 mAhg$^{-1}$, reaching about 97% of the theoretical value[56]. The achieved theoretical specific energy density is 380 Whkg$^{-1}$, *i.e.*, a value comparable with that offered by commercially-available batteries[5]. Nevertheless, the use of ultralight, high capacity graphene nanoflakes anode allows to estimate a practical energy density of about 190 Whkg$^{-1}$, *i.e.* a value exceeding (~25-60%) that of current lithium ion battery technology[5].

The stability of electrode materials is another key point for battery cycling. Fig. 4d shows that the battery has a very good cycling behaviour, operating at 1C rate for more than 80 charge-discharge cycles with high Coulombic efficiency. The latter increases from 89% at the first cycle (this low initial irreversibility is due to SEI film formation at the LiFePO$_4$ cathode side), to reach about 99.5% at its steady-state operation.

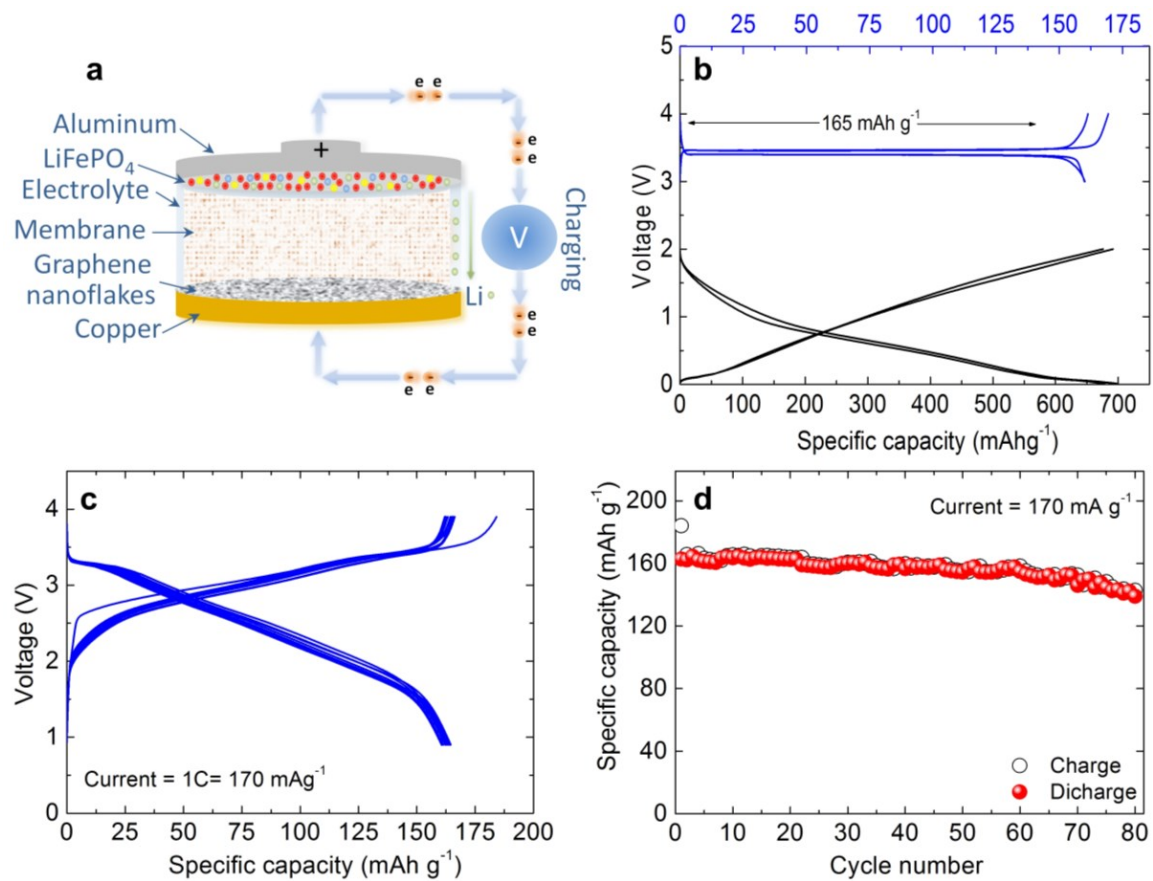

**Figure 4**. **Electrochemical test of a graphene nanoflakes/lithium iron phosphate battery. a**, Schematic of graphene/lithium iron phosphate battery. **b**, Charge–discharge voltage profiles of the single electrodes, *i.e.* the graphene nanoflakes anode (black curve) and the LiFePO$_4$ cathode (blue curve) as reported versus lithium. Current rate 170 mAg$^{-1}$ (LiFePO$_4$) and 700 mAg$^{-1}$ (graphene nanoflakes). **c**, Voltage profile of the graphene/LiFePO$_4$ full battery. **d**, Specific capacity versus cycle number of the battery. Electrolyte: LP30. Cycling rate 1C (170 mA g$^{-1}$ vs. LiFePO$_4$). Voltage limits 0.9-3.9 V. Temperature 25 $^o$C.

This battery, based on Cu-supported graphene nanoflakes, exhibits a much better performances in terms of specific capacity, cyclability and Coulombic efficiency, with respect to the one previously reported and based on RGO as anode material[32]. The excellent performance in terms of cycling life and rate capability of this battery, to the best of our knowledge so far rarely reported, confirms the potentiality of graphene nanoflakes as innovative electrode material for the progress of the lithium-based, energy storage systems.

**Conclusions**

In summary, we have designed a high-performance anode produced via drop casting ink of graphene nanoflakes with on-demand controlled morphological properties. The Cu-supported graphene nanoflakes anode has large Li ion uptake (LiC$_2$ stochiometry) coupled with excellent electrical (sheet resistance ~0.5 Ω/□) and electrochemical properties, *i.e.* specific capacity exceeding 1,500 mAg$^{-1}$ at a current rate of 100 mAg$^{-1}$ and long cycle life (over 150 charge-discharge cycles). These high performance electrodes has been exploited, coupled with lithium iron phosphate cathode, for the realization of the first Li-ion battery based on graphene nanoflakes. This novel battery shows excellent performances in terms of reversible capacity (165 mAhg$^{-1}$), cycle life (>80 cycles) and Coloumbic efficiency (~100%). The use of inks of graphene nanoflakes with controlled geometry represents a new, promising and cost effective direction for the development of next generation Li-ion batteries for portable electronics, electric vehicles and grid-scale applications. This successful materials design for anodes could also be extended to cathode systems as well as exploited in other energy storage devices such as Li-air batteries and supercapacitors.

**Methods**

**Preparation of graphene ink.** The graphene ink used in this work was prepared dispersing 200 mg of Graphite flakes (Sigma Aldrich Ltd.) in 20 ml of N-Methyl2Pyrrolidone (NMP, Sigma Aldrich Ltd.)[34,36]. The initial dispersion was then ultrasonicated (Branson 3510) for 6 hours and subsequently ultracentrifuged using a SW-41 swinging bucket rotor in a Beckman-Coulter Optima XPN ultracentrifuge at 25000rpm (~77000g) for 30 mins. After ultracentrifugation, the supernatant was extracted by pipetting.

**Characterization of the graphene ink.** The concentration of graphitic flakes in the as-prepared ink was determined from α at 660 nm, as described in Refs. [34,36]. Absorption measurements were carried out with a Jasco V-550 UV-Vis. For Raman measurements, the graphene ink was diluted with N-Methyl2Pyrrolidone and drop-casted onto a Si wafer with 300 nm thermally grown $SiO_2$. Raman measurements were carried out with a Renishaw 1000 at 532 nm and a 100X objective, with an incident power of ~1mW. The D, G and 2D peaks were fitted with Lorentzian functions. The as-prepared ink was also drop casted at room temperature (RT) onto carbon coated copper TEM grids (300 mesh) and rinsed with DI water. TEM images were taken with a Zeiss Libra 120 transmission electron microscope, operated at 120kV and equipped with an in-column omega filter. All the images and the diffraction patterns were energy filtered with a 15eV slit on the zero loss peak.

**Preparation of the electrodes.** Graphene ink was drop casted on a polycrystalline Cu substrate at 140°C at room condition. 6.25 ml of ink was deposited on each Cu substrate. To remove traces of undesired solvent residual, the electrode was annealed at 400 °C in UHV for 3 hours.

**Characterization of the electrodes.** Electrical characterization of the Cu-supported graphene nanoflakes electrode was carried out with a Jandel station RM3 with multi height probe combined with the X, Y, θ ball bearing coordinate stage with 0-25 mm × 0.01 μm screws and rotary stage of the Microposition Probe head, 100 nm titanium tips arranged in a straight line 1mm apart, combined with a Keithley 2100 digital multimeter. The measurement accuracy was verified against a 12.93 Ω/□ indium Tin oxide on glass reference (Jandel Engineering Ltd., tested against a NIST traceable sample). The SEM measurements were performed with a field-emission Zeiss Auriga 405 instrument with beam energy of 10 KeV, working distance of 3.3mm and 1.0nm nominal resolution. The XPS measurements were performed with un-monochromatized Al $K_\alpha$ radiation ($h\nu$=1486.7 eV). Photoelectrons were analyzed with a CLAM-2 VG hemispherical electron analyzer, 100mm average radius, pass energy of 100eV. The BE was calibrated with respect to a clean Au sample in electrical contact with the Cu-supported graphene nanoflakes electrode by measuring the Au-$4f_{7/2}$ core level at 84.0 eV. All measurements were performed at RT in UHV conditions (pressure $10^{-10}$ mbar). Exposure to lithium was carried out in-situ in UHV conditions, by means of commercial SAES Getters dispensers. The C-1s and Li-1s core-levels are fitted with Voigt functions (Lorentzian-Gaussian peaks).

**Assembly and electrochemical characterization of the batteries**. The LiFePO4 electrode film was prepared by blending the active material (80%), super P carbon (10%, Timcal, electron conductor) and polyvinylidene fluoride (10%, PVdF6020, binder, Sigma Aldrich Ltd.) in NMP (solvent, Sigma Aldrich Ltd.); the slurry was then drop cast on aluminum foil, and finally dried overnight under vacuum at 110°C. The active material loading was about 1 mg cm$^{-2}$. Prior to full lithium ion cell assembling, the Cu-supported graphene nanoflakes electrode was partially pre-lithiated by a surface treatment[57]. This was performed by placing the electrode in direct contact with a Li foil wet by the electrolyte solution (*i.e.*, LP30, EC:DMC 1:1, LiPF6 1M Merck) for 30 min.

The galvanostatic cycling tests on lithium half cells were carried out with a Maccor battery tester using Swagelok type cells prepared by coupling the electrode under test with a lithium foil counter electrode in an EC;DMC, 1:1, LiPF$_6$, 1M electrolyte soaked in a glass fibre separator (Whatman). The cycling tests of the Graphene-Cu electrode in half lithium cell were performed at various current densities (*i.e.* 100, 400 and 700 mA g$^{-1}$) in the 0.01V – 2.0V voltage range. The full Cu-supported graphene nanoflakes/LiFePO$_4$ battery was evaluated by galvanostatic cycling in Swagelok type cell formed by coupling the pre-treated anode[57] with the cathode in the EC;DMC, 1:1, LiPF6, 1M electrolyte with a glass fibre separator (Whatman). The battery was slightly cathode limited and was cycled at 1C rate (170 mAg$^{-1}$ based on the cathode weight) in the 0.9V-3.9V voltage range.

**Acknowledgements**

We acknowledge C. Coletti and F. Carillo for useful discussions and the Graphene Flagship (contract no. CNECT-ICT-604391) for financial support.

**Appendix A: Exfoliation and sorting of graphitic flakes**

Graphene nanoflake inks are prepared via low-power ultrasonication of graphite in N-Methyl2Pyrrolidone (NMP). The choice of graphite exfoliation in NMP is set by the need of having the best contact at the inter-flake junctions upon deposition on the electrode, which could be affected by the surfactant coverage in aqueous solutions[37,58]. During the ultrasonication process, the strong hydrodynamic shear-force, created by the propagation of cavitons[59], *i.e.*, the creation and subsequent collapse of bubbles or voids in liquids due to pressure fluctuations[59], induces exfoliation of the graphitic flakes[36]. However, the exfoliation process produces a heterogeneous dispersion of thin/thick and small/large graphitic flakes[15]. After exfoliation, the solvent-graphene interaction needs to balance the inter-flakes attractive forces that cause re-aggregation.

The sorting of small lateral size graphene nanoflakes is carried out by SBS. During the SBS process, graphitic flakes dispersed in a solvent under centrifugal fields are subjected to three forces, see Fig. 5. The centrifugal force, $F_s=m\omega^2 r$, equal to the product of the mass of the graphitic flake (m), the square of the angular velocity ($\omega$), and the distance (r) from the rotational axis[38,60]. The buoyant force, $F_b=-m_0\omega^2 r$, equal to the product of the weight of fluid displaced ($m_0$) by the graphitic flakes (Archimedes' principle)[61], the square of $\omega$, and r.

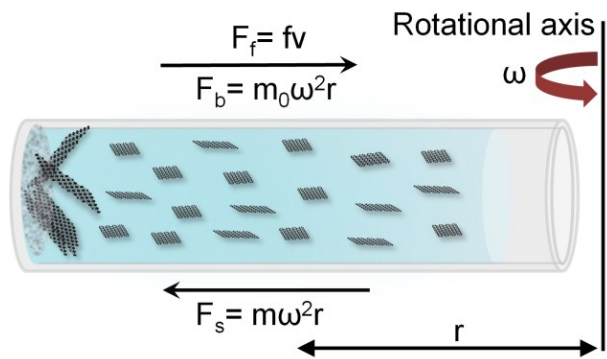

**Figure 5**: Forces acting on graphene flakes inside an ultracentrifuge tube during ultracentrifugation with a swinging bucket rotor. The buoyant force $F_b$ and the frictional force, $F_f$ act in opposite directions to the centrifugal force $F_s$, and thus opposite to sedimentation.

The frictional force[60], $F_f=-fv$, where f is the frictional coefficient due to the motion through the solvent towards the bottom of the ultracentrifuge tube and *v* is the sedimentation velocity. In general, a particle of known volume and density in a medium of constant density will be accelerated under a centrifugal field, until the net force on the particle equals the force resisting its motion through the medium[60,61]. *f* depends on mass and shape of the particles[60,62] and increases as the particle geometry moves away from a spherical shape, which means that large or elongated particles experience more frictional drag than spherical ones having the same mass[63].

The rate of sedimentation of a graphitic flake in a centrifugal field is described by the Svedberg equation[38]:

$$s = v/\omega^2 r = m(1-\dot{\upsilon}\rho)/f \quad (1)$$

where s is the sedimentation coefficient, the time needed for flakes to sediment, commonly reported in Svedberg (S) unit (1S corresponds to 10–13 sec.)[38], $\dot{\upsilon}$ is the partial specific volume (the volume that each gram of the solute occupies in solution) and $\rho$ is the density of the solvent. s depends on the morphological properties of the particle and is proportional to the buoyant effective molar weight of the particle, while it is inversely proportional to f [38]. As reported in Eq. 1, the sedimentation of graphitic flakes depends on the frictional coefficient and mass[15]. Thick and large flakes, having larger mass, sediment faster with respect to thin and small flakes (having smaller mass), which are thus retained in dispersion.

**Appendix B: Characterization of graphene nanoflakes ink**

We use optical absorption spectroscopy (OAS) in order to evaluate the concentration (c) of graphitic material in the ink. Fig. 6 plots the OAS of the ink prepared via SBS. The UV absorption peak at ~266nm is attributed to inter-band electronic transitions from the unoccupied $\pi^*$ states at the M point of the Brillouin zone[64,65]. The asymmetry of the UV peak, with a high-wavelength tail, is attributed to excitonic effects[65,66]. Using the experimentally derived absorption coefficient of 1390 Lg$^{-1}$m$^{-1}$ at 660 nm [34,36] we estimate c ~40 mg/l.

Fig. 7 plots a low-resolution transmission electron microscopy (TEM) bright field image revealing a large quantity of flakes deposited on the TEM grid. The sample is formed by flakes having variable dimensions mostly in the range ~30–100 nm.

Electron diffraction collected on flake aggregates shows polycrystalline rings demonstrating that the flakes are crystalline. All the rings can be indexed as h,k,-h-k,0 reflections of an hexagonal lattice with a=0.244(1)nm, in agreement with the graphene structure[67].

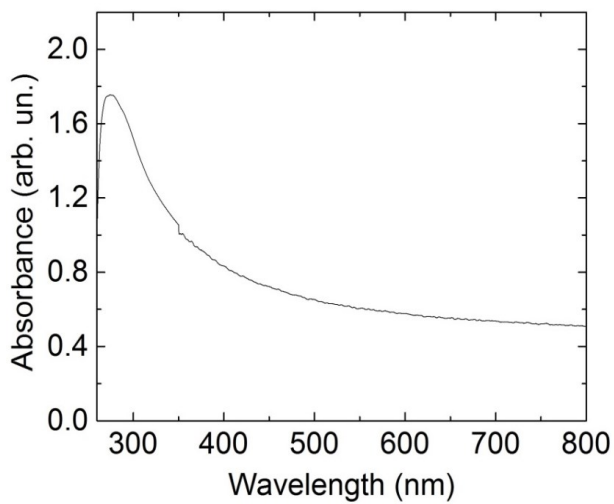

**Figure 6**: Room temperature (RT) absorption spectrum of the SBS graphene ink.

Raman spectroscopy is a fast and non-destructive technique to identify number of layers, doping, defects, disorder, chemical modifications and edges of graphitic flakes[39,68,69]. In a typical Raman spectrum of graphene, the G peak corresponds to the $E_{2g}$ phonon at the Brillouin zone centre[39,68,69]. The D peak is due to the breathing modes of sp$^2$ rings and requires a defect for its activation by double resonance[39,42]. The 2D peak is the second order of the D peak[39]. This is a single peak in monolayer graphene, whereas it splits in four structures in bi-layer graphene, reflecting the evolution of the band structure[39]. The 2D peak is always seen, even when no D peak is present, since no defects are required for the activation of two phonons with the same momentum, one backscattered from the other[39]. Double resonance can also happen as an intra-valley process, *i.e.* connecting two points belonging to the same cone around K or K' [68]. This process gives rise to the D' peak, while the 2D' is the second order of the D'.

Statistical analysis of the micro-Raman spectra (Fig. 8) shows that the 2D peak is at Pos(2D)~2691cm$^{-1}$ (Fig. 8a), while the FWHM(2D) varies from 60 to 95cm$^{-1}$ with a peak at ~75cm$^{-}$ [1](Fig. 8b)) and I(2D)/I(G) varies from 0.75 to 1.05 (Fig. 8c)). This is consistent with the samples being a combination of single layer (SLG) and few-layer graphene (FLG) flakes. The Raman spectra show significant D and D' peaks intensity, with an average intensity ratio I(D)/I(G) ~1.50 (see Fig. 8d) and I(D')/I(G) ~0.35.

This is attributed to the edges of our nanometer flakes[40], rather than to structural defects on the basal plane of SLG and FLG flakes. This observation is supported by the analysis of I(D)/I(G) (Fig. 8d), FWHM(G) (Fig. 8e) and Pos(G) (Fig. 8f). Indeed, combining I(D)/I(G) with FWHM(G) allows us to discriminate between disorder localized at the edges and disorder in the bulk. In the latter case, a higher I(D)/I(G) would correspond to higher FWHM(G). I(D)/I(G) and FWHM(G) are not correlated, as shown in the inset to Fig. 1e, an indication that the major contribution to the D peak comes from the sample edges. Moreover, in the high-defect concentration regime FWHM(G) and FWHM(D') become broader and eventually merge into a single band[68].

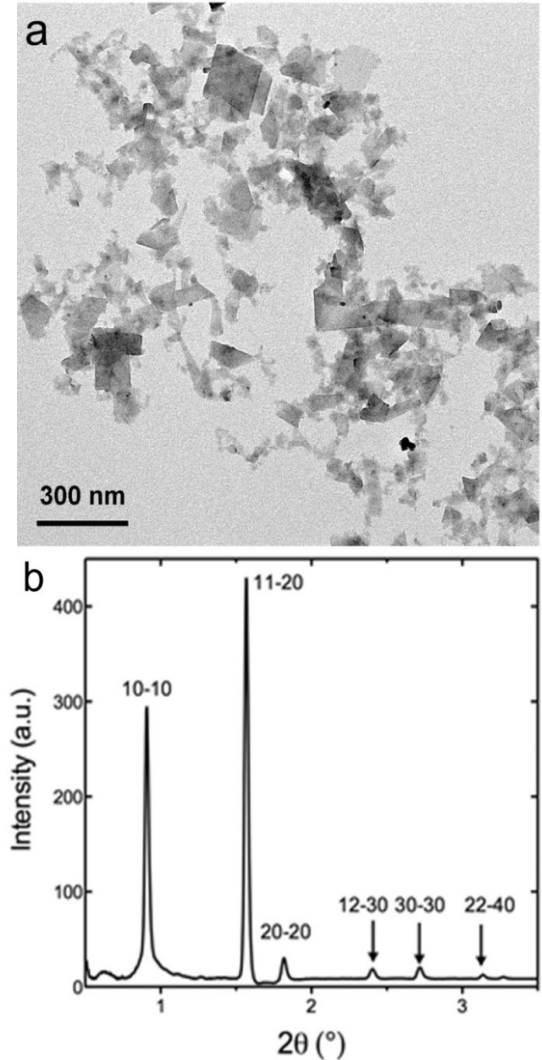

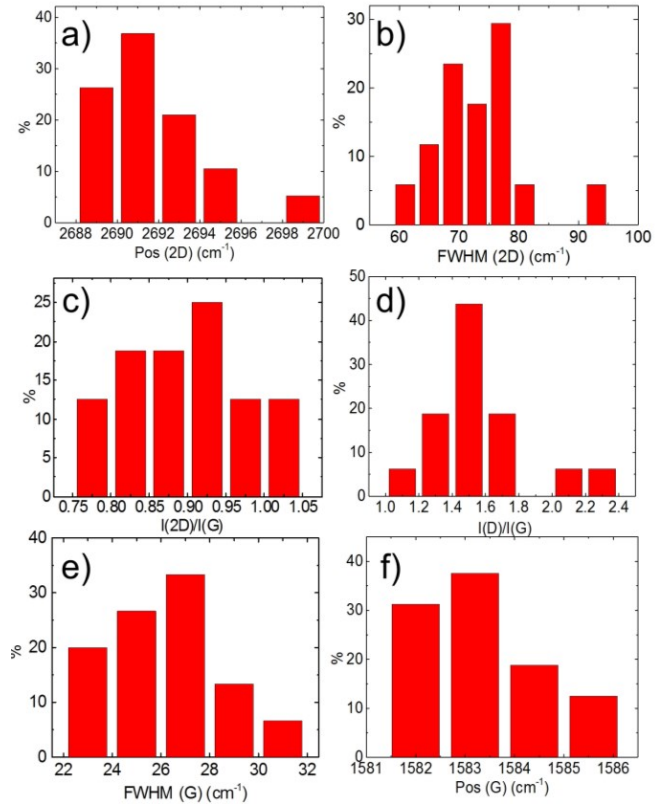

**Figure 7**: (a) Bright field TEM images of aggregates of graphene flakes at low magnification. (b) Plot of the diffracted intensity vs. the 2θ (angle between the incident direction and the direction where the scattering is observed) scattering angle obtained by a radial integration of the electron diffraction pattern collected on an area of 2 micron in diameter. The peaks are indexed according to the graphene structure[67].

**Figure 8**: Statistic on Raman measurements carried out at a laser excitation of 532nm. Histograms showing the distribution of a) Pos(2D), b) FWHM(2D), c) I(2D)/I(G), d) I(D)/I(G), e) FWHM(G), and f) Pos(G).

## Appendix c: Characterization of graphene nanoflakes electrode

The *Rs* of the anode electrode is measured with a Jandel station (Model RM3000) with 4-Probe head, 100 μm titanium tips arranged in a straight line 1mm apart, combined with a digital multimeter. The electrical current (I) that flows through the outer probes induces a voltage drop (V) between the two inner probes. When the probes spacing is equal and small compared to the size of the sample (electrode) being measured, the voltage difference between the inner probes may be defined as70:

$$V = (IRs)/\pi \ln(2) \quad (2)$$

leading to:

$$Rs = (V/I) \times (\pi/\ln(2)) \approx V/I \times 4.53 \quad (3)$$

Sheet resistance measurements are carried avoiding the sample edges in order to verify the approximation of the four probe method70. The measurement accuracy is verified against a 12.93Ω/□ indium Tin oxide on glass reference (Jandel Engineering Ltd., Se. No 74703 tested against a NIST traceable sample).

We carried out 10 measurements on the Cu-supported graphene nanoflakes electrode. The average value of *Rs* is ~0.5 Ω/□. This extremely low *Rs* value is attributed to the crystallinity of the graphene nanoflakes. The low *Rs* is also favoured by the good electric contact with the Cu substrate.

Fig. 9 compares the statistical Raman analysis of flakes deposited from the ink, with the measurements carried out on the electrode. Fig. 9a and Fig. 9b compare the Pos(2D) and FWHM(2D) distributions. The data show that the electrode (cyan bars) has Pos(2D) upshifted (~10cm$^{-1}$) and FWHM(2D) slightly larger (4cm$^{-1}$) with respect to that of the graphene flakes in the ink. However, both Pos(2D) and FWHM(2D) distributions of the Raman spectrum taken on the electrode remains distinctly different from that of graphite[42], an indication that graphene flakes are electronically decoupled from each other. Fig. 10a plots the C 1s core-level of the thermally treated (400°C) Cu-supported graphene nanoflakes electrode. The spectrum is dominated by a well-defined peak at 284.1 eV binding energy (BE), associated to high-purity graphene[44] with a weak tail due to carbon-oxygen bonds[45], *i.e.* C-O, C=O and O-C=O [41].

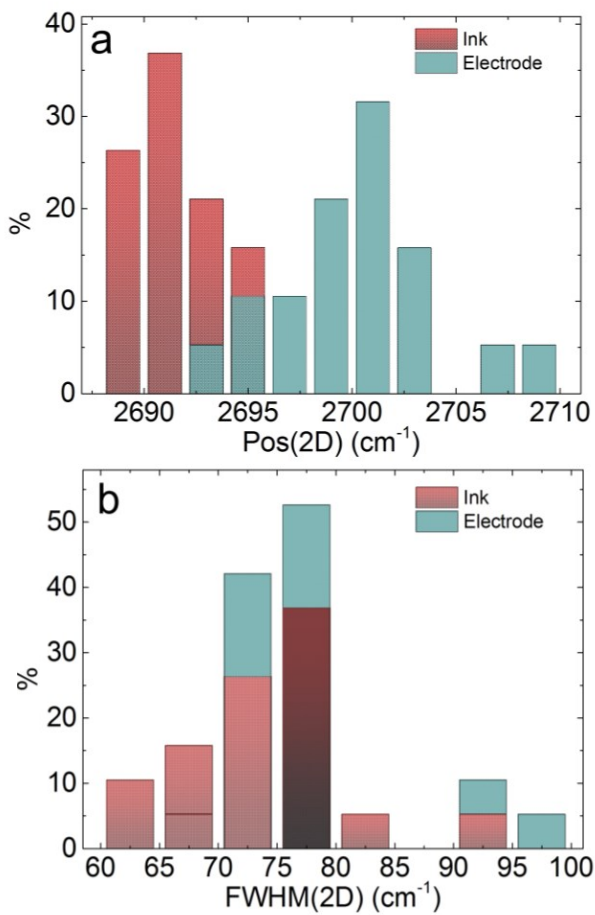

**Figure 9**: Statistic on Raman measurements carried out at 532nm. Histograms showing the distribution of a) Pos(2D) and b) FWHM(2D) of the ink (light wine) and the Cu-supported graphene nanoflakes electrode (dark cyan).

By fitting the experimental data with with Voigt functions (Lorentzian-Gaussian peaks) taking into consideration the intrinsic linewidth (Lorentzian contribution) and the experimental resolution (Gaussian contribution), we estimate that the 284.1 eV C-C peak accounts for more than 83%, being the carbon-oxygen bonds associated to contamination from solvent residual and/or drop casting deposition method at room conditions.

Fig. 10b shows the evolution of the XPS signal of the C 1s peak of the Cu-supported graphene nanoflakes electrode, upon increasing exposure time to Li at RT in controlled evaporation UHV conditions. The C 1s XPS signal associated to graphene upshifts up to a value of 284.5 eV BE and further Li deposition does not influence the lineshape and the binding energy, suggesting the saturation of Li uptake. Fig. 10c plots the Li1s XPS signal (peak at 55.7 eV BE) associated with uptake of lithium in the Cu-supported graphene nanoflakes electrode. The Li intensity increases as a function of exposure time and the signal associated to Li ions coordinated with C atoms saturates at 60' exposure.

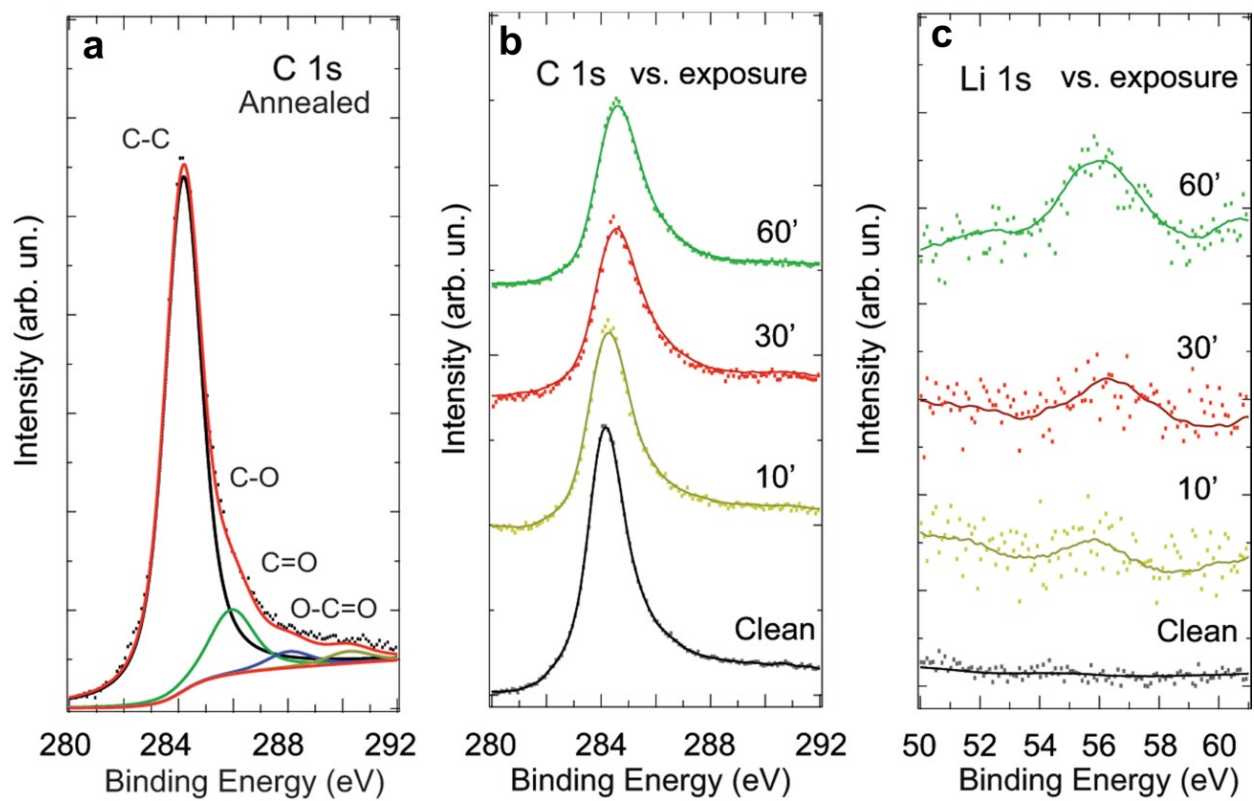

**Figure 10**: (a) XPS data at the C-1s core-level of the Cu-supported graphene nanoflakes electrode after annealing in UHV at 400 °C. The contributions of C-C, C-O, C=O and O-C=O are determined by fitting the curve with Voigt functions71. XPS signals of the (b) C-1s and (c) Li-1s core-levels as a function of exposure time of the Cu-supported graphene nanoflakes electrode to Li in UHV. Data are taken at RT. Spectra are vertically shifted for clarity.

## References

[1] US Department of Energy: http://www1.eere.energy.gov/vehiclesandfuels/electric_vehicles/10_year_goal.html

[2] Yoshino, A.; Sanechika, K. & Nakajima, T. Japanese Patent 1989293,1985.

[3] Burke, A., Miller, M. & Zhao, H. Ultracapacitors in hybrid vehicle applications: testing of new high

3 power devices and prospects for increased energy density. *Research Report - UCD-ITS-RR-12-06*, (2012).
4 Tarascon, J. M. & Armand, M. Issues and challenges facing rechargeable lithium batteries. *Nature* **414**, 359-367 (2001).
5 Scrosati, B. & Garche, J. Lithium batteries: Status, prospects and future *J. Power Sources*, **195** 2419-2430 (2010).
6 Goodenough, J. B. & Park, K-S. The Li-Ion Rechargeable Battery: A Perspective. *J. Am. Chem. Soc.* **135**, 1167–1176 (2013).
7 Armand, M. & Tarascon, J. M. Building better batteries. *Nature* **451,** 652–657 (2008).
8 Wilson, A., Way, B., Dahn, J. & Van Buuren, T. Nanodispersed silicon in pregraphitic carbons. *J. Appl. Phys.*, **77**, 2363-2369 (1995).
9 Yu, Y. Gu, L. Dhanabalan, A. Chen C.-H. & Wang, C. *Electrochim. Acta* **54**, 7227-7230 (2009).
10 Paek, S. M., Yoo, E. & Honma, I. Enhanced cyclic performance and lithium storage capacity of $SnO_2$/graphene nanoporous electrodes with three-dimensionally delaminated flexible structure. *Nano Lett*. **9**, 72–75 (2009).
11 Dimov, N., Kugino, S. & Yoshio, M. Carbon-coated silicon as anode material for lithium ion batteries: advantages and limitations, *Electrochimica Acta* **48**, 1579-1587 (2003).
12 Stoller, M. D., Park, S., Zhu, Y., An, J. & Ruoff, R. S. Graphene-based ultracapacitors. *Nano Lett.* **8**, 3498-3502 (2008).
13 Geim, A. K. & Novoselov, K. S. The rise of graphene. *Nature Mater*. **6**, 183-189 (2007).
14 Lee, C., Wei, X., Kysar, J. W. & Hone, J. Measurement of the elastic properties and intrinsic strength of monolayer graphene. *Science* **321**, 385-388 (2008).
15 Bonaccorso, F. *et al.* Production and processing of graphene and 2d crystals. *Mater. Today* **15**, 564-589 (2012).
16 Bonaccorso, F. *et al*. Graphene, related two dimensional crystals, and hybrid systems for energy conversion and storage, submitted (2014).
17 Xu, C. *et al.* Graphene-based electrodes for electrochemical energy storage. *Energy Environ. Sci.* **6**, 1388-1414 (2013).
18 Li, X. *et al*. Large-Area Synthesis of High-Quality and Uniform Graphene Films on Copper Foils. *Science* **324**, 1312-1314 (2009).
19 Pollack, E. *et al.* The interaction of $Li^+$ with single-layer and few-layer graphene. *Nano Lett.* **10**, 3386–3388 (2010).
20 Lian, P. *et al*. Large reversible capacity of high quality graphene sheets as an anode material for lithium-ion batteries. *Electrochim. Acta* **55**, 3909–3914 (2010).
21 Yoo, E. *et al*. Large reversible Li storage of graphene nanosheet families for use in rechargeable lithium ion batteries. *Nano Lett.* **8**, 2277-2282 (2008).
22 Wu, Z. S. *et al*. Graphene Anchored with $Co_3O_4$ Nanoparticles as anode of lithium ion batteries with enhanced reversible capacity and cyclic performance. *ACS Nano* **4**, 3187-3194 (2010).
23 Wang, H. *et al.* $Mn_3O_4$–graphene hybrid as a high-capacity anode material for lithium ion batteries. *J. Am. Chem. Soc*. **132**, 13978-13980 (2010).
24 Zhou, G. *et al*. Graphene-wrapped $Fe_3O_4$ anode material with improved reversible capacity and cyclic stability for lithium ion batteries. *Chem. Mater*. 22, 5306–5313 (2010).
25 Evanoff, K., Magasinski, A., Yang, J. & Yushin, G. Nanosilicon-coated graphene granules as anodes for Li-Ion batteries. *Adv. Energy Mater.* **1**, 495-498 (2011).
26 Yang, S., Feng, X., Ivanovici, S. & Müllen, S. Fabrication of graphene-encapsulated oxide nanoparticles: towards high-performance anode materials for lithium storage. *Angew. Chem. Int. Ed*. **49**, 8408-8411 (2010).
27 Wang, H., *et al.* $LiMn_{1-x}Fe_xPO_4$ nanorods grown on graphene sheets for ultrahigh-rate-performance lithium ion batteries. *Angew. Chem. Int. Ed.* **50**, 7364-7368 (2011).
28 Kim, H., *et al.*, Graphene-based hybrid electrode material for high-power lithium-ion batteries. *J. Electrochem. Soc*. **158**, A930 (2011).
29 Bak, S.-M. *et al.* Spinel $LiMn_2O_4$/reduced graphene oxide hybrid for high rate lithium ion batteries. *J. Mater. Chem.* **21**, 17309-17315 (2011).
30 Wang, H. *et al.* Graphene-wrapped sulfur particles as a rechargeable lithium-sulfur battery cathode material with high capacity and cycling stability. *Nano Lett.* **11**, 2644-2647 (2011).
31 Mattevi, C. *et al*. Evolution of electrical, chemical, and structural properties of transparent and conducting chemically derived graphene thin films. *Adv. Funct. Mater*. **19**, 2577-2583 (2009).
32 Vargas, O., Caballero, A., Morales, J., Elia, G. A., Scrosati, B. & Hassoun, J. Electrochemical

performance of a graphene nanosheets anode in a high voltage lithium-ion cell. *Phys. Chem. Chem. Phys.* **15**, 20444-20446 (2013).

33  Uthaisar, C. & Barone, V. Edge effects on the characteristics of Li diffusion in graphene. *Nano Lett.* **10**, 2838-2842 (2010).

34  Torrisi, F. *et al.* Inkjet-printed graphene electronics. *ACS Nano* **6**, 2992-3006 (2012).

35  Wang, G., Shen, X., Yao, J. & Park, J. Graphene nanosheets for enhanced lithium storage in lithium ion batteries. *Carbon* **47**, 2049-2053 (2009).

36  Hernandez, Y. *et al*. High-yield production of graphene by liquid-phase exfoliation of graphite. *Nature Nanotech*. **3**, 563-568 (2008).

37  Maragó, O. M. *et al.* Brownian motion of graphene. *ACS Nano* **4**, 7515-7523 (2010).

38  Svedberg, T. & Pedersen, K. O. The Ultracentrifuge, Oxford Univ. Press, London **1940**.

39  Ferrari, A. C. *et al*. Raman spectrum of graphene and graphene layers. *Phys. Rev. Lett.,* **97**, 187401 (2006).

40  Casiraghi, C. *et al.* Raman spectroscopy of graphene edges. *Nano Lett*. **9**, 1433-1441 (2009).

41  Nam Han, N. *et al.* Improved heat dissipation in gallium nitride light-emitting diodes with embedded graphene oxide pattern. *Nature Comm*. **4**, 1452 (2013).

42  Tuinstra, F. & Koenig, J. L. Raman spectrum of graphite. *J. Chem. Phys.* **53**, 1126-1130 (1970).

43  Siegbahn, K. Electron spectroscopy for chemical analysis. Springer US, 1973.

44  Lizzit, S. *et al.* Band dispersion in the deep 1s core level of graphene. *Nature Phys*. **6**, 345-349 (2010).

45  Mordkovich, V. Z. Synthesis and XPS investigation of superdense lithium-graphite intercalation compound, $LiC_2$. *Synth. Met*. **80**, 243-247 (1996).

46  Yeh, J. J. & Lindau, I. Atomic subshell photoionization cross sections and asymmetry parameters: $1 < Z < 103$. *At. Data Nucl. Data Tables* **32**, 1-155 (1985).

47  Nalimova, V. A., Guérard, D., Lelaurain, M. & Fateev, O.V. X-Ray investigation of highly saturated Li-graphite intercalation compound. *Carbon* **33**, 177-181 (1995).

48  Rabii, S. & Guérard, D. Stability of superdense lithium graphite compounds, *J. Phys. And Chem. Of Solids* **69**, 1165-1167 (2008).

49  Kar, T., Pattanayak, J. & Scheiner, S. Insertion of lithium ions into carbon nanotubes: an ab initio study. *J. Phys. Chem. A* **105**, 10397–1040 (2001).

50  Kashedikar, N. A. & Maier, J. Lithium storage in carbon nanostructures. *Adv. Mater.* **21**, 2664-2680 (2009).

51  Dahn, J. R., Zheng, T., Liu, Y. & Xue, J. S. Mechanisms for lithium insertion in carbonaceous materials. *Science* **270**, 590-593 (1995).

52  Fong, R. von Sacken, U. & Dahn, J. R. Studies of lithium intercalation into carbons using nonaqueous electrochemical cells. *J. Electrochem. Soc.* **137**, 2009-2013 (1990).

53  Lin, J. *et al.* Graphene Nanoribbon and Nanostructured $SnO_2$ Composite Anodes for Lithium Ion Batteries. *ACS Nano* **7,** 6001–6006 (2013).

54  Bhardwaj, T., Antic, A., Pavan, B., Barone, V. & Fahlman, B. D. Enhanced electrochemical lithium storage by graphene nanoribbons. *J. Am. Chem. Soc*. **132**, 12556–12558 (2010).

55  Li, L., Raji, A-R. O. & Tour, J. M. Graphene-wrapped $MnO_2$–graphene nanoribbons as anode materials for high-performance lithium ion batteries. *Adv. Mater*. **25**, 6298–6302 (2013).

56  Croce, F. *et al*. A novel concept for the synthesis of an improved $LiFePO_4$ lithium battery cathode. *Electrochem. Solid-State Lett.* **5**, A47-A50 (2002).

57  Hassoun, J., Lee, K.-S., Sun, Y.-K. & Scrosati, B. An advanced lithium ion battery based on high performance electrode materials. *J. Am. Chem. Soc*. **133**, 3139–3143 (2011).

58  Lotya M. *et al.* Liquid phase production of graphene by exfoliation of graphite in surfactant/water solutions. *J. Am. Chem. Soc.* **131**, 3611-3620 (2009).

59  Mason T. J. Sonochemistry; Oxford University: New York, Chapter 1. 1999.

60  Williams, J. W., Van Holde, K. E., Baldwin, R. L. & Fujita, H. The Theory of sedimentation analysis. *Chem. Rev*. **58**, 715 (1958).

61  Piazza, R., Buzzaccaro, S., Secchi, E. & Parola, A. What buoyancy really is. A Generalized Archimedes principle for sedimentation and ultracentrifugation. *Soft Matter* **8**, 7112-7115 (2012).

62  Schuck, P. Size-distribution analysis of proteins by analytical ultracentrifugation: strategies and application to model systems. *Biophys. J*. **78**, 1096-1111 (2000).

63  Wilson, K. & Walker, J. Principles and techniques of biochemistry and molecular biology, Cambridge Univ. Press, 1993.

[64] Greenaway, D. L., Harbeke, G., Bassan, F. & Tosatti, E. Anisotropy of the optical constants and the band structure of graphite *Phys. Rev.*, **178**, 1340-1348 (1969).

[65] Kravets, V. G. *et al.*, Spectroscopic ellipsometry of graphene and an exciton-shifted van Hove peak in absorption. *Phys. Rev. B*, 2010, **81**, 155413.

[66] Yang, L.; Deslippe, J.; Park, C. H.; Cohen,M. L.; Louie, S. G. Excitonic effects on the optical response of graphene and bilayer graphene. *Phys. Rev. Lett.* **103**, 186802 (2009).

[67] Meyer, J. C., Geim, A. K., Katsnelson, M. I., Novoselov, K. S. Booth, T. J. & Roth, S. The structure of suspended graphene sheets. *Nature* **446**, 60-63 (2007).

[68] Ferrari, A. C. & Robertson, J. Interpretation of Raman spectra of disordered and amorphous carbon. *Phys. Rev. B*, **61**, 14095 (2000).

[69] Ferrari, A. C. & Basko, D. Raman spectroscopy as a versatile tool for studying the properties of graphene. *Nature Nanotech.* **8**, 235-246 (2013).

[70] Smits, F. M. Measurement of sheet resisitivities with the four-point probe. *Bell Sys. Tech. Jour.*, **37**, 711-718 (1958).

[71] Temme, N. M. "Voigt function", NIST Handbook of Mathematical Functions, Cambridge University Press, (2010).